\documentclass[preprint,showpacs,superscriptaddress,nofootinbib]{revtex4}
  \usepackage{graphicx,subfigure}
      \usepackage{color}
    
\begin{document}
  \title{Constraints on hard spectator scattering and annihilation corrections in $B_{u,d}$ ${\to}$ $PV$ decays within QCD factorization}
  \author{Junfeng Sun}
  \affiliation{Institute of Particle and Nuclear Physics,
              Henan Normal University, Xinxiang 453007, China}
  \author{Qin Chang}
  \email{changqin@htu.edu.cn}
  \affiliation{Institute of Particle and Nuclear Physics,
              Henan Normal University, Xinxiang 453007, China}
  \affiliation{State Key Laboratory of Theoretical Physics, Institute of Theoretical Physics,
        Chinese Academy of Sciences, China}
  \author{Xiaohui Hu}
  \affiliation{Institute of Particle and Nuclear Physics,
              Henan Normal University, Xinxiang 453007, China}
  \author{Yueling Yang}
  \affiliation{Institute of Particle and Nuclear Physics,
              Henan Normal University, Xinxiang 453007, China}
  \begin{abstract}
  In this paper, we investigate the contributions of hard spectator scattering 
  and annihilation in $B$ ${\to}$ $PV$ decays within the QCD factorization framework.
  With available experimental data on $B$ ${\to}$ ${\pi}K^{\ast}$,
  ${\rho}K$, ${\pi}{\rho}$ and $K{\phi}$ decays, comprehensive 
  $\chi^2$ analyses of the parameters $X_{A,H}^{i,f}$$({\rho}_{A,H}^{i,f}, {\phi}_{A,H}^{i,f})$ are performed,
  where $X_{A}^{f}$ ($X_{A}^{i}$) and $X_{H}$ are used to
  parameterize the endpoint divergences of the (non)factorizable
  annihilation and hard spectator scattering amplitudes, respectively.
  Based on $\chi^2$ analyses, it is observed that
  (1) The topology-dependent parameterization scheme is feasible
  for $B$ ${\to}$ $PV$ decays; 
 (2) At the current accuracy of experimental measurements and theoretical evaluations,
  $X_{H}$ $=$ $X_{A}^{i}$ is allowed by $B$ ${\to}$ $PV$ decays, but $X_{H}$ $\neq$ $X_{A}^{f}$ at $68\%$ C. L.;
  (3) With the simplification $X_{H}$ $=$ $X_{A}^{i}$, parameters $X_{A}^{f}$ and $X_{A}^{i}$ should be treated individually.
  The above-described findings are very similar to those obtained from $B$ ${\to}$ $PP$ decays. Numerically, for $B$ ${\to}$ $PV$ decays, we obtain $({\rho}_{A,H}^{i},{\phi}_{A,H}^{i}[^{\circ}]) =(2.87^{+0.66}_{-1.95}, -145^{+14}_{-21})$ and $({\rho}_{A}^{f},{\phi}_{A}^{f}[^{\circ}]) = (0.91^{+0.12}_{-0.13}, -37^{+10}_{-9})$ at $68\%$ C. L.. 
  With the best-fit values, most of the theoretical results are in good agreement with the experimental data within errors. However, significant corrections to the color-suppressed tree amplitude $\alpha_2$ related to a large ${\rho}_{H}$ result in the wrong sign for $A^{dir}_{CP}(B^{-}{\to}{\pi}^{0}K^{{\ast}-})$ compared with the most recent BABAR data, which presents a new obstacle in solving ``${\pi}{\pi}$'' and ``${\pi} K$" puzzles through $\alpha_2$. A crosscheck with measurements at Belle (or Belle II) and LHCb, which offer higher precision, is urgently expected to confirm or refute such possible mismatch.
  \end{abstract}
  \maketitle

  Nonleptonic decays of hadrons containing a heavy quark play an important 
  role in testing the Standard Model (SM) picture of the $CP$ violation mechanism 
  in flavor physics, improving our understanding of nonperturbative and 
  perturbative QCD and exploring new physics beyond the SM.
  For charmless $B$ meson decays, experimental studies have been successfully
  carrying out at $B$ factories ( BABAR and Belle) and 
  Tevatron ( CDF and D0) in the past and will be continued
  by running LHCb and upgrading Belle II experiments.
  These experiments provide highly fertile ground for theoretical studies 
  and have yielded many exciting and important results, such as measurements
  of pure annihilation $B_{s}$ ${\to}$ ${\pi}^{+}{\pi}^{-}$ and $B_{d}$
  ${\to}$ $K^{+}K^{-}$ decays reported recently by CDF, LHCb and Belle 
  \cite{CDFanni,LHCbanni,PRD87.031103}, which may suggest
  the existence of unexpected large annihilation contributions and have attracted much 
  attention, for instance, Refs. \cite{xiao1,Gronau,zhu1,zhu2,chang1,chang2}.
  
  Theoretically, to calculate the hadronic matrix elements of hadronic $B$ weak decays,
  some approaches, including QCD factorization (QCDF)
  \cite{Beneke1}, perturbative QCD~(pQCD) \cite{KLS}
  and soft-collinear effective theory~(SCET) \cite{scet},
  have been fully developed and extensively employed in recent years.
 Even though the annihilation contributions are formally power suppressed in the
heavy quark limit, they may be numerically important for realistic hadronic $B$ decays, particularly
for pure annihilation processes and direct $CP$ asymmetries.
Unfortunately, in the collinear factorization approximation, the calculation of annihilation corrections always suffers from end-point divergence. In the pQCD approach, such divergence is regulated by introducing the parton transverse momentum $k_{T}$ and the Sudakov factor at the expense of modeling the additional $k_{T}$ dependence of meson wave functions, and large complex annihilation corrections are presented~\cite{LvCD}.  In the SCET approach, such divergence is removed by separating the physics at
  different momentum scales and using zero-bin subtraction
  to avoid double counting the soft degrees of freedom \cite{msf,scetAnni}; thus,
  the annihilation diagrams are factorable but real to the
  leading power term of ${\cal O}(\alpha_{s}(m_{b})\Lambda_{QCD}/m_{b})$.
The absence of strong phases from SCET's annihilation
 amplitudes differs with the pQCD's estimation
  and the QCDF expectation \cite{Beneke2}.
  
Within the QCDF framework,   
to estimate the annihilation amplitudes and regulate the endpoint
  divergency, the logarithmically divergent integral is usually parameterized
  in a model-independent manner \cite{Beneke2} and explicitly
  expressed as
  \begin{equation}
  {\int}_{0}^{1} \frac{dx}{x}\;{\to}\; 
  X_{A}=(1+{\rho}_{A} e^{i{\phi}_{A}}){\ln}\frac{m_{b}}{{\Lambda}_{h}}
  \label{XA},
  \end{equation}
  with the typical scale ${\Lambda}_{h}$ $=$ 0.5 GeV. 
  Moreover, a similar endpoint singularity also appears in the hard spectator 
  scattering (HSS) contributions of higher twist distribution amplitudes that
  are also formally power suppressed but chirally enhanced; therefore, a similar 
  parameterization ansatz is used to cope with HSS endpoint divergency,
  and quantity $X_{H}$ (${\rho}_{H}$, ${\phi}_{H}$), similar to
  the definition of Eq.(\ref{XA}), is introduced. 
  As discussed in Ref. \cite{Beneke2},  
  $X_{H,A}$ ${\sim}$ ${\ln}(m_{b}/{\Lambda}_{h})$ is expected
  because the effects of the intrinsic transverse 
  momentum and off-shellness of partons would be to modify  
  $x$ ${\to}$ $x$ $+$ ${\epsilon}$ with 
  ${\epsilon}$ ${\sim}$ ${\cal O}({\Lambda}_{\rm QCD}/m_{b})$
  in the denominator of Eq.(\ref{XA}).
  The factor $(1+{\rho}e^{i{\phi}})$ summarizes the remaining
  unknown nonperturbative contributions, where ${\phi}$, which is related to the strong phase,
   is important for direct $CP$ asymmetries. 
   In such a parameterization scheme, even though the predictive power of QCDF is partly weakened due to the incalculable parameters $\rho$ and ${\phi}$ that are introduced, it also provides a feasible way to evaluate the effects and the behavior of annihilation and HSS corrections from a phenomenological view point, which is helpful for understanding and exploring possible underlying mechanisms.
   
  Although the magnitude of and constraints on parameter ${\rho}$
  are utterly unknown based on the first principles of QCD dynamics for now, 
  an excessively large value of ${\rho}$ would significantly enhance the subleading $1/m_{b}$ contributions,
  and hence, a conservative choice of ${\rho}_{A}$ ${\sim}$ 1 has typically been used
  in previous phenomenological studies \cite{Beneke2,Cheng3,Cheng1,Cheng2}. In practice, 
  different values of $({\rho}_{A},{\phi}_{A})$ chosen according to various B meson decay types~($PP$, $PV$, $VP$ and $VV$) have been used to fit experimental data ~\cite{Beneke2,Cheng2}.
  However, with the favored ``Scenario S4", in which
  ${\rho}_{A}$ ${\simeq}$ $1$ and ${\phi}_{A}$ ${\simeq}$ $-55^{\circ}$ \cite{Beneke2} for $B$ ${\to}$ $PP$ decay,
  the QCDF prediction 
  ${\cal B}(B_{s}{\to}{\pi}^{+}{\pi}^{-})$ $=$ 
  $(0.26^{+0.00+0.10}_{-0.00-0.09}){\times}10^{-6}$
  \cite{Cheng2} is about $3.4{\sigma}$ less than the 
  experimental data $(0.73{\pm}0.14){\times}10^{-6}$ \cite{HFAG}. 

  Motivated by this possible mismatch, detailed analyses have been 
  performed within the QCDF framework \cite{zhu1,zhu2,chang1,chang2}. 
  In Refs. \cite{zhu1,zhu2}, a ``new treatment" for endpoint parameters is presented in which the flavor
  dependence of the annihilation parameter $X_A$ on the initial
  states should be carefully considered, and hence, $X_A$ is divided
  into two independent parameters $X_A^i$ and $X_A^f$,
 which are responsible for parameterizing the endpoint divergences 
  of nonfactorizable and factorizable annihilation topologies, respectively.
  Following the proposal of Refs. \cite{zhu1,zhu2} and combining 
  available experimental data for $B_{u,d,s}$ ${\to}$ ${\pi}K$,
  ${\pi}{\pi}$ and $K\bar{K}$ decays, the comprehensive 
  $\chi^2$ analyses of $X^{i,f}_{A}$ and $X_{H}$ in $B\to PP$ decays
  were performed in Refs. \cite{chang2,chang3}. 
  It was found that 
\begin{itemize}
\item Theoretically, there is neither a compulsory constraint
  nor a priori reason for both $X^{i}_{A}$ $=$ $X^{f}_{A}$ $=$ $X_{A}$
  and $X_{A}$ being universal for all hadronic $B$ decays;
  Phenomenologically, it is required by available measurements 
  regarding $B\to PP$ decays that $X^{i}_{A}$ and $X^{f}_{A}$ should be 
  treated individually; in addition, the simplification $X_{H}$  $=$ $X_{A}^{i}$ is allowed by data, which effectively reduces the number of unknown variables, but  $X_{H}$  $\neq$ $X_{A}^{f}$  (see scenario III in Ref. \cite{chang2} for detail);
\item The effect of flavor symmetry breaking on parameter $X_{A}^{i,f}$
  is tiny and negligible for the moment due to large experimental
  errors and theoretical uncertainties;
\item A slightly large ${\rho}_{H}$ ${\sim}$ $3$ with 
  ${\phi}_{H}$ ${\sim}$ $-105^{\circ}$ and a relatively small 
  inverse moment parameter ${\lambda}_{B}$ ${\sim}$ 200 MeV
  for $B$ meson wave functions are required to enhance the 
  color-suppressed coefficients ${\alpha}_{2}$ with a large 
  strong phase, which is important in accommodating all available
  observables of $B_{u,d,s}$ ${\to}$ ${\pi}K$, ${\pi}{\pi}$ and 
  $K\bar{K}$ decays simultaneously, even the 
  so-called ``${\pi}K$'' and ``${\pi}{\pi}$'' puzzles
  (see Refs. \cite{chang2,chang3} for detail); 
\item Numerically, in the most simplified scenario in which $X_{H}$  $=$ $X_{A}^{i}$ is assumed, combining the constraints from $B_{u,d,s}$ 
  ${\to}$ ${\pi}K$, ${\pi}{\pi}$ and $K\bar{K}$ decays, two solutions responsible for $B\to PP$ decays are obtained \cite{chang3},
    \begin{eqnarray}
   {\rm Solution\ A:} & & 
    \left\{ \begin{array}{l}
  ({\rho}_{A,H}^{i},{\phi}_{A,H}^{i}[^{\circ}]) = 
  (2.98^{+1.12}_{-0.86},-105^{+34}_{-24}), 
    \\
  ({\rho}_{A}^{f},{\phi}_{A}^{f}[^{\circ}]) = 
  (1.18^{+0.20}_{-0.23},-40^{+11}_{-~8}),
    \\
   {\lambda}_{B} =
  0.19^{+0.09}_{-0.04}\,{\rm GeV};
    \end{array} \right.
    \label{PPSA}
    \\
   {\rm Solution\ B:} & &
    \left\{ \begin{array}{l}
  ({\rho}_{A,H}^{i},{\phi}_{A,H}^{i}[^{\circ}]) = 
 (2.97^{+1.19}_{-0.90},-105^{+32}_{-24}),
    \\
  ({\rho}_{A}^{f},{\phi}_{A}^{f}[^{\circ}]) =
  (2.80^{+0.25}_{-0.21},165^{+4}_{-3}),
   \\
  {\lambda}_{B} =
  0.19^{+0.10}_{-0.04}\,{\rm GeV},
   \end{array} \right.
   \label{PPSB}
   \end{eqnarray}
which yield similar HSS and annihilation contributions.
\end{itemize}
 
  In recent years, many measurements of $B$ ${\to}$ $PV$ decays 
  have been performed anew at higher precision \cite{HFAG}. Thus, with the available experimental 
  data, it is worth reexamining the agreement between QCDF's
  predictions and experimental data on $B$ ${\to}$ $PV$ decays,
  investigating the effects of HSS and annihilation contributions, and further testing whether the aforementioned findings regarding $B$ ${\to}$ $PP$ decays still persist in $B$ ${\to}$ $PV$ decays. 
  In this paper, we would like to extend our previous studies on $B$
  ${\to}$ $PP$ decays \cite{chang2,chang3} to $B$ ${\to}$ $PV$ 
  decays with the same ${\chi}^2$ fit method and similar treatment 
  of annihilation and HSS parameters; the details of the statistical
  ${\chi}^2$ approach can be found in the appendix of Refs.
  \cite{chang2,Vernazza}. 
  
  For $B$ ${\to}$ $PV$ decays, the decay amplitudes and 
  relevant formulae have been clearly listed in Ref. \cite{Beneke2}.
  The parameters $X_{A}^{i,f}$ under discussion appear in the basic
  building blocks of annihilation amplitudes, which can be explicitly written as follows \cite{Beneke2}:
  \begin{eqnarray}
  & & 
  A_{1}^{i}\ {\simeq} -A_{2}^{i}\
  {\simeq}\ 6{\pi}{\alpha}_{s} \Big[
  3\Big(X_{A}^{i}-4+\frac{{\pi}^{2}}{3}\Big)
  +r_{\chi}^{M_{1}}r_{\chi}^{M_{2}}
   \Big( (X_{A}^{i})^{2}-2X_{A}^{i} \Big) \Big]
   \label{ampAnni1}, \\
  & &
   A_{3}^{i}\ {\simeq}\ 6{\pi}{\alpha}_{s} \Big[
   -3r_{\chi}^{M_{1}} \Big( (X_{A}^{i})^{2}-2X_{A}^{i}-\frac{{\pi}^{2}}{3}-4 \Big)
   + r_{\chi}^{M_{2}} \Big( (X_{A}^{i})^{2}-2X_{A}^{i}-\frac{{\pi}^{2}}{3} \Big) \Big]
   \label{ampAnni2}, \\
  & &
  A_{1}^{f} = A_{2}^{f} =0
  \label{ampAnni3}, \\
  & &
  A_{3}^{f}\ {\simeq}\ 6{\pi}{\alpha}_{s} \Big[
    3r_{\chi}^{M_{1}} ( 2X_{A}^{f}-1) ( 2-X_{A}^{f} )
   -  r_{\chi}^{M_{2}} \Big( 2 (X_{A}^{f})^{2}- X_{A}^{f} \Big) \Big]
  \label{ampAnni4},
  \end{eqnarray}
  for the $VP$ final state, where the superscript $f$ ($i$) in $A^{f(i)}_{k}$ corresponds to (non)factorizable 
  annihilation topologies. For the $PV$ final state, one must simply exchange $r_{\chi}^{M_{1}}$
  ${\leftrightarrow}$ $r_{\chi}^{M_{2}}$ and change the sign of $A_{3}^{f}$. Further explanation and information on QCDF's annihilation amplitudes can be found in Ref. \cite{Beneke2}.
  
  Before entering further discussion, we would like to note the following: 
  (1) In previous studies, the annihilation parameters were assumed to be process-dependent \cite{Beneke2,Cheng3,Cheng1,Cheng2} where
  (${\rho}^{PV}_{A}$, ${\phi}^{PV}_{A}$) and
  (${\rho}^{VP}_{A}$, ${\phi}^{VP}_{A}$) were introduced to
  describe nonleptonic $B$ decay into the final states 
  $PV$ and $VP$ decays, respectively;
  sometimes, additional values of (${\rho}_{A}$, ${\phi}_{A}$)
  for $B$ ${\to}$ $K{\phi}$ decays \cite{Cheng1} were required.
  In our analysis, parameters (${\rho}^{i}_{A}$, ${\phi}^{i}_{A}$) 
  and (${\rho}^{f}_{A}$, ${\phi}^{f}_{A}$) are topology-dependent.
  (2) As discussed in Refs. \cite{zhu1,zhu2,chang2}, parameters
  $X_{A}^{f}$(${\rho}^{f}_{A}$, ${\phi}^{f}_{A}$) are assumed
  to be universal for factorizable annihilation amplitudes and free
  of flavor-symmetry-breaking effects because
  they are not associated with the wave function of initial $B$ mesons,
  and the approximations of the asymptotic light cone distribution amplitudes of the final 
  states are used.
  (3) The wave function of $B$ mesons is involved in the 
  calculation of nonfactorizable annihilation amplitudes. 
  Generally, the momentum fraction of light $u$, $d$ quarks
  in $B_{u,d}$ mesons should be different from that of the 
  spectator $s$ quark in $B_{s}$ meson.
  The flavor-symmetry-breaking effects might be embodied in 
  parameters $X_{A}^{i}$(${\rho}^{i}_{A}$, ${\phi}^{i}_{A}$).
  In this paper, only $B_{u,d}$ ${\to}$ $PV$ decays are
  considered~(most $B_{s}$ ${\to}$ $PV$ decays have not been measured), and the
  isospin symmetry is assumed to be held.
  (4) Unlike in the case of $B$ ${\to}$ $PP$ decays, 
  in which both final states are pseudoscalar mesons,
  the wave functions of the vector mesons are also required to
  evaluate the hadronic matrix elements of $B$ ${\to}$ $PV$ 
  decays. Therefore, following the treatment of annihilation parameters presented in 
  Refs. \cite{Beneke2,Cheng3,Cheng1,Cheng2},
  the parameters $X_{A}^{i,f}$(${\rho}^{i,f}_{A}$, ${\phi}^{i,f}_{A}$) for $B$ ${\to}$ 
  $PV$ decays are generally different from those for $B$ 
  ${\to}$ $PP$ decays.
  
  As is well known, for the $b$ ${\to}$ $s$ transition, the tree 
  contributions are strongly suppressed by the CKM factor 
  ${\vert}V_{us}^{\ast}V_{ub}{\vert}$ ${\sim}$ ${\cal O}({\lambda}^{4})$,
  whereas the penguin contributions are proportional to the CKM factor
  ${\vert}V_{cs}^{\ast}V_{cb}{\vert}$ ${\sim}$ ${\cal O}({\lambda}^{2})$
  \cite{PDG14}.
  In addition, the nonfactorizable contributions between vertex and 
  HSS corrections largely cancel each other out \cite{Beneke2}. 
  Therefore, the weak annihilation amplitudes are important for the 
  $b$ ${\to}$ $s$ nonleptonic $B$ decays.
  Large annihilation contributions are derived from the coefficient $b_{3}$,
  because $b_{3}$ is proportional to the CKM factor 
  ${\vert}V_{cs}^{\ast}V_{cb}{\vert}$ and sensitive to the annihilation
  building block $A_{3}^{f}$, which is always accompanied by $N_{c}C_{6}$.
  Hence, it is expected that precise observables of $b$ ${\to}$ $s$
  nonleptonic $B$ decays could introduce stringent restrictions on parameters
  $X_{A}^{f}$(${\rho}^{f}_{A}$, ${\phi}^{f}_{A}$).  
  
  For the $b$ ${\to}$ $d$ transition, the tree contributions are dominant
  if they exist, whereas the penguin contributions are suppressed due to the 
  cancellation between the CKM factor $V_{ud}^{\ast}V_{ub}$ and 
  $V_{cd}^{\ast}V_{cb}$ \cite{PDG14}.
  Large annihilation contributions are derived from the coefficient $b_{1,2}$,
  which is always accompanied by large Wilson coefficients $C_{1,2}$.
  For color-suppressed tree-dominated hadronic $B$ decays,
  the contributions of HSS and factorizable annihilation corrections
  are particularly important, for example, the resolution of the so-called
  ``${\pi}{\pi}$'' puzzle \cite{chang2}.  
  Therefore, severe restrictions on parameters $X_{A,H}^{i}$ could be derived from many precise
  observables of the $b$ ${\to}$ $d$ nonleptonic $B$ decays.
  
  The decay modes considered in this paper include the penguin-dominated $B$ ${\to}$ ${\pi}K^{\ast}$, ${\rho}K$ decays
  induced by the $b$ ${\to}$ $s\bar{q}q$ ($q$ $=$ $u$, $d$)
  transition, the penguin-dominated $B$ ${\to}$ ${\phi}K$ 
  decays induced by the $b$ ${\to}$ $s\bar{s}s$ transition,
  the tree-dominated $B$ ${\to}$ ${\pi}{\rho}$ decays 
  induced by the $b$ ${\to}$ $d\bar{q}q$ transition, and the
  penguin- and annihilation-dominated $B$ ${\to}$ $KK^{\ast}$
  decays induced by the $b$ ${\to}$ $d\bar{s}s$ transition. 
For the observables of the above-mentioned decay modes, the available experimental data are summarized in the ``Exp." columns of Tables \ref{brcp}, \ref{mcp} and \ref{cpp}, in which most of data are the averaged results given by HFAG~\cite{HFAG}, except for the branching fractions and direct CP asymmetries of  $B^{-}$ ${\to}$ ${\pi}^{-}\bar{K}^{{\ast}0}$, ${\pi}^{0}K^{{\ast}-}$ and $\bar{K}^{0}{\rho}^{-}$ decays. Recently, using the full dataset of $470.9\pm2.8$ million $B\bar{B}$ events, the BABAR collaboration reported the latest results from an analysis of $B^+\to K^0\pi^+\pi^0$~(and the combined results from this and previous BABAR analyses)~\cite{babarpik}
   \begin{eqnarray}
  B^{-}{\to}{\pi}^{-}\bar{K}^{{\ast}0}:& & 
    \left\{ \begin{array}{l}
 {\cal B}[\times 10^{-6}]=14.6\pm2.4\pm1.4^{+0.3}_{-0.4}\quad(11.6\pm0.5\pm1.1),     \\
  A_{CP}^{dir}[\%]=-12\pm21\pm8^{+0}_{-11}\quad(2.5\pm5.0\pm1.6);
      \end{array} \right.
    \label{bb1}
    \\
B^{-}{\to}{\pi}^{0}K^{{\ast}-}:& &
    \left\{ \begin{array}{l}
 {\cal B}[\times 10^{-6}]=9.2\pm1.3\pm0.6^{+0.3}_{-0.5} \quad(8.8\pm1.1\pm0.6),   \\
  A_{CP}^{dir}[\%]=-52\pm14\pm4^{+4}_{-2}\quad(-39\pm12\pm3);
     \end{array} \right.
   \label{bb2}
       \\
B^{-}{\to}\bar{K}^{0}{\rho}^{-}:& &
    \left\{ \begin{array}{l}
 {\cal B}[\times 10^{-6}]=9.4\pm1.6\pm1.1^{+0.0}_{-2.6},    \\
  A_{CP}^{dir}[\%]=21\pm19\pm7^{+23}_{-19},
     \end{array} \right.
   \label{bb3}
   \end{eqnarray}
in which, in particular, the first evidence of a CP asymmetry of $B^{-}{\to}{\pi}^{0}K^{{\ast}-}$ is observed at the $3.4\sigma$ significance level. In our following analysis, such (combined) results for $B^{-}{\to}{\pi}^{0}K^{{\ast}-}$ and $\bar{K}^{0}{\rho}^{-}$ decays in Eqs.~(\ref{bb2}) and~(\ref{bb3}) are used. For $B^{-}{\to}{\pi}^{-}\bar{K}^{{\ast}0}$ decay, its branching fractions and direct CP asymmetry are also measured by Belle collaboration~\cite{bellepik}; therefore, we adopt the weighted averages of observables, which are presented in Table \ref{brcp}.

The data listed in Tables \ref{brcp}, \ref{mcp} and \ref{cpp}
  demonstrate that the first three sets of decay modes are well
  measured; therefore, experimental data of these decay modes
  are used in our fitting. In addition, the theoretical inputs are summarized in the Appendix. 
  Our following analyses and fitting are divided into three cases for different purposes. 

 %%%%%%%%%%%%%%%%%%%%%%%%%%%%%%%%%%%%%%%%
 \begin{figure}[t]
 \subfigure[]{\includegraphics[width=5.0cm]{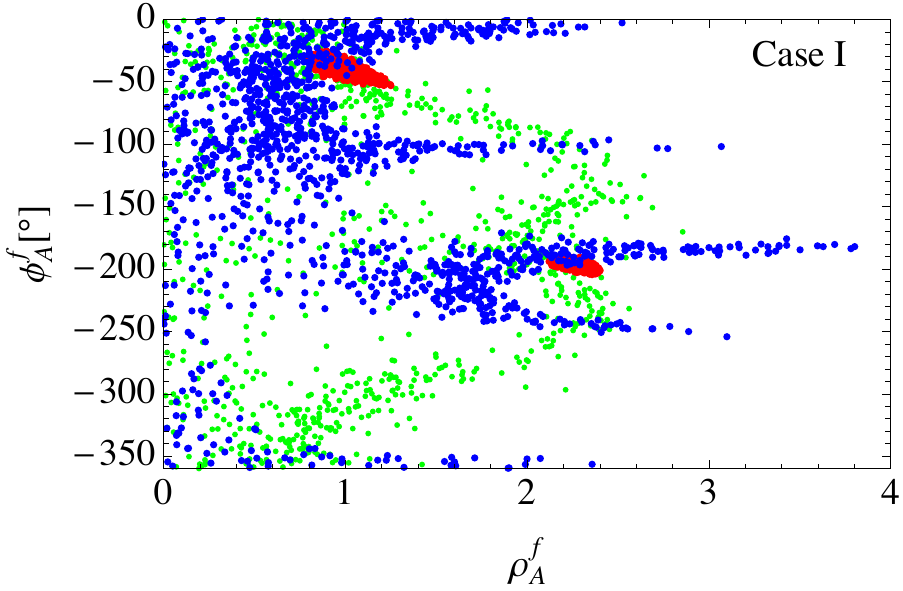}} \
 \subfigure[]{\includegraphics[width=5.0cm]{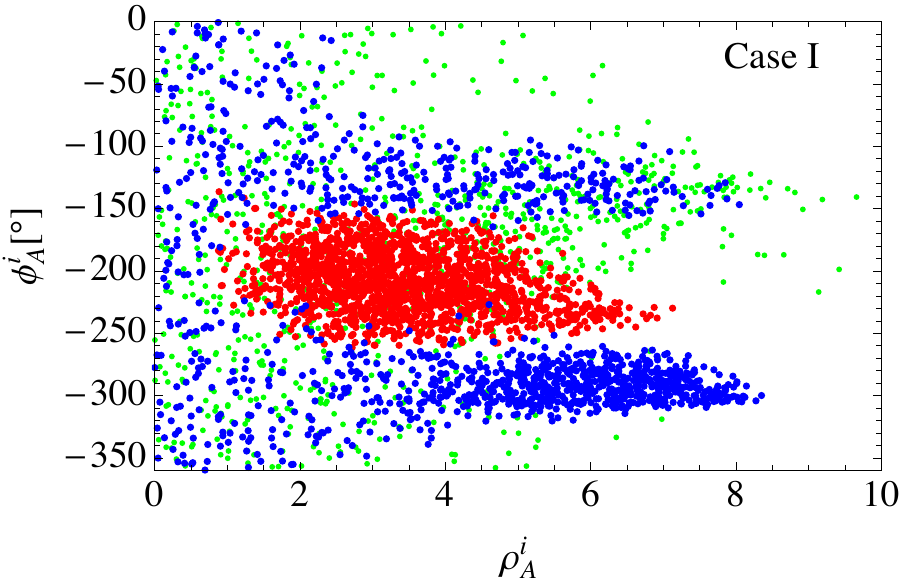}} \
 \subfigure[]{\includegraphics[width=5.0cm]{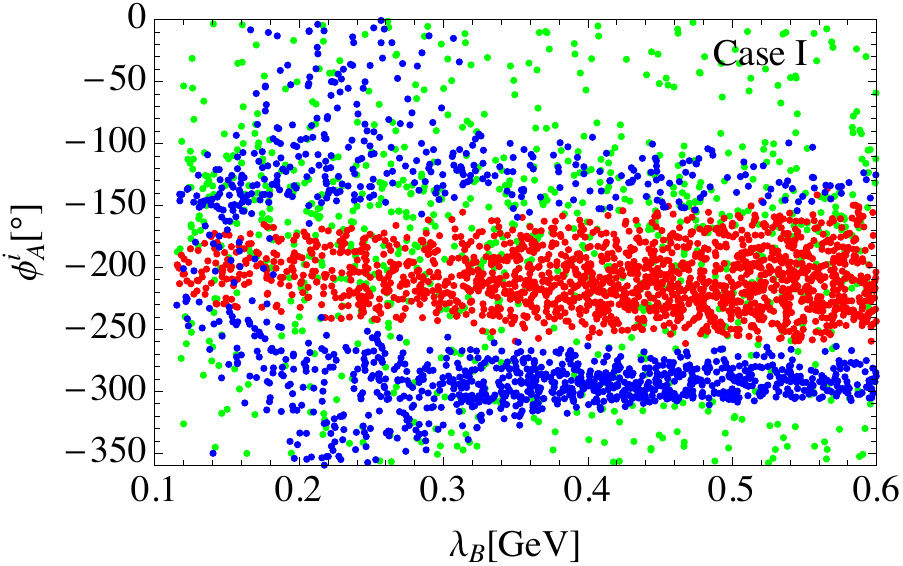}}
 \caption{The allowed regions of parameters (${\rho}^{i,f}_{A}$, ${\phi}^{i,f}_{A}$) and 
  ${\lambda}_{B}$ at $68\%$ C.L. with the constraints from $B$ ${\to}$ ${\pi}K^{\ast}$, ${\rho}K$ 
 decays (red), $B$ ${\to}$ ${\pi}{\rho}$ decays (blue), and $B$ ${\to}$
 ${\phi}K$ decays (green), respectively. }
 \label{figCaseI}
 \end{figure}
 %%%%%%%%%%%%%%%%%%%%%%%%%%%%%%%%%%%%%%%%

  (1) 
  For case I, five parameters, (${\rho}^{i,f}_{A}$, ${\phi}^{i,f}_{A}$) and 
  ${\lambda}_{B}$, are treated as free parameters, and the simplification
  $X_{H}$ $=$ $X_{A}^{i}$, which is allowed in $B$ ${\to}$ $PP$ 
  decays \cite{chang2}, is assumed. Moreover, the constraints from $B$ ${\to}$ ${\pi}K^{\ast}$, ${\rho}K$ 
 decays, $B$ ${\to}$ ${\pi}{\rho}$ decays, and $B$ ${\to}$
 ${\phi}K$ decays are considered separately. 
  The fitted results are shown in Fig. \ref{figCaseI}. 
  
  Fig. \ref{figCaseI} (a) clearly shows that
  parameters (${\rho}^{f}_{A}$, ${\phi}^{f}_{A}$) are strictly 
  bound into two separate compact regions (red points) around 
  $(0.9,-40^{\circ})$ and $(2.2,-200^{\circ})$ 
  by the constraints from $B$ ${\to}$ ${\pi}K^{\ast}$, ${\rho}K$ 
  decays, which is similar to the case for $B$ ${\to}$ $PP$ decays 
  (see Eq.(\ref{PPSA}) and Eq.(\ref{PPSB})).
  Moreover, these two regions overlap with the 
  blue and green dotted regions, which implies 
  that the two solutions of (${\rho}^{f}_{A}$, ${\phi}^{f}_{A}$)
  are also allowed by $B$ ${\to}$ ${\pi}{\rho}$, ${\phi}K$ decays. 
  
  As shown in Fig. \ref{figCaseI} (b), under the constraints from $B$ ${\to}$ ${\pi}{\rho}$ decays, the parameters (${\rho}^{i}_{A}$, ${\phi}^{i}_{A}$)
  are loosely restricted into two wide bands (blue points) around
  ${\phi}^{i}_{A}$ ${\sim}$ $-130^{\circ}$ and ${\sim}$ $-300^{\circ}$ because
  the experimental precision of the observables, especially the direct $CP$ asymmetries, on $B$ ${\to}$ ${\pi}{\rho}$
  decays is still very rough. Under the constraints from $B$ ${\to}$ ${\pi}K^{\ast}$ and ${\rho}K$  decays,   (${\rho}^{i}_{A}$, ${\phi}^{i}_{A}$) are restricted around ${\phi}^{i}_{A}$ ${\sim}$ $-200^{\circ}$~(red points) and overlap partly with the blue pointed region, which implies that the allowed spaces of (${\rho}^{i}_{A}$, ${\phi}^{i}_{A}$) would be seriously shrunken under the combined constraints.
    
  From Fig. \ref{figCaseI} (c), parameter ${\lambda}_{B}$ cannot be 
  determined exclusively, although an additional phenomenological 
  condition 115 MeV ${\leq}$ ${\lambda}_{B}$ ${\leq}$ 600 MeV
  is imposed during our fit based on the studies of Refs. \cite{Beneke2,
  PRD.80.054024,PRD.80.111105,NPB.832.109,Beneke5,PLB.718.1014}.
  In principle, parameter ${\lambda}_{B}$ is only related to the $B$ wave
  function and independent of any decay modes.
  Therefore, in our following analyses, the result ${\lambda}_{B}$ $=$ $0.19^{+0.09}_{-0.04}$ GeV fitted from $B$ ${\to}$ $PP$ decays \cite{chang3} will be adopted. 

  %%%%%%%%%%%%%%%%%%%%%%%%%%%%%%%%%%%%%%%%
  \begin{figure}[htb]
  \includegraphics[width=7cm]{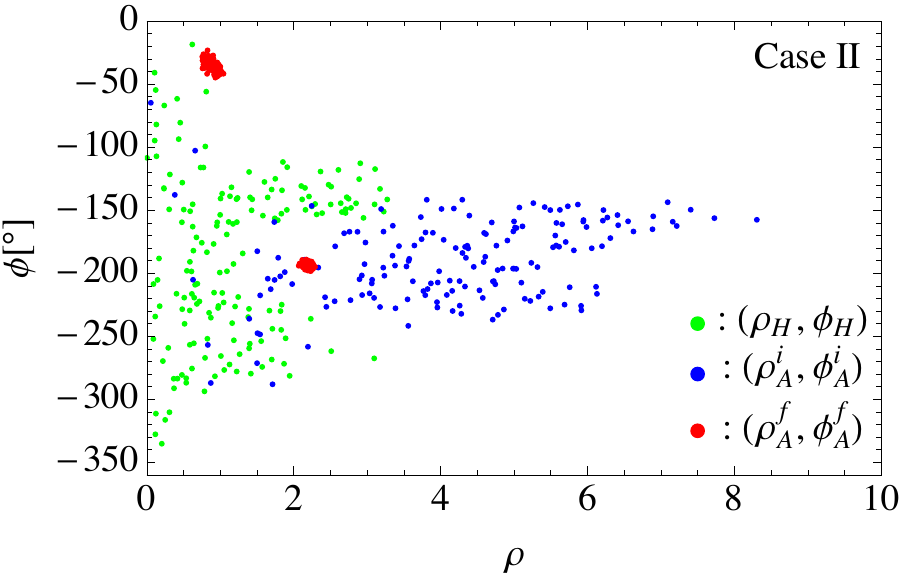}
  \caption{The allowed regions of parameters (${\rho}^{i,f}_{A,H}$, ${\phi}^{i,f}_{A,H}$) at 68\% C.L.}
  \label{figCaseII}
  \end{figure}
 %%%%%%%%%%%%%%%%%%%%%%%%%%%%%%%%%%%%%%%%
 
 (2) 
 For Case II, to determine whether the simplification $X_{H}$ $=$ $X_{A}^{i}$
 is valid for $B$ ${\to}$ $PV$ decays, both (${\rho}_{H}$, ${\phi}_{H}$) and 
 (${\rho}^{i,f}_{A}$, ${\phi}^{i,f}_{A}$) are treated as free parameters.
 Combining all available constraints from $B$ ${\to}$ ${\pi}K^{\ast}$, ${\rho}K$,
 ${\pi}{\rho}$, ${\phi}K$ decays, the allowed parameter spaces at 
 68\% C.L. are shown in Fig. \ref{figCaseII}. 

 Fig. \ref{figCaseII} clearly shows that 
 (i) 
Similarly to Case I, two solutions of (${\rho}^{f}_{A}$, ${\phi}^{f}_{A}$) with very 
 small uncertainties (red points) are obtained, which are denoted ``solution A" for ${\phi}^{f}_{A}\sim -40^{\circ}$ and ``solution B" for ${\phi}^{f}_{A}\sim -200^{\circ}$ for convenience;
Meanwhile, the spaces of (${\rho}^{i}_{A}$, ${\phi}^{i}_{A}$) are still hardly well bounded (blue points) as in Case I;
 (ii)
 The allowed spaces of (${\rho}^{f}_{A}$, ${\phi}^{f}_{A}$) are
 small and tight, whereas those of (${\rho}^{i}_{A}$, ${\phi}^{i}_{A}$)
 are big and loose; thus, they generally differ from each other.
This finding implies that $X_{A}^{f}$ and $X_{A}^{i}$ may
 be treated individually, as in the case for $B$ ${\to}$
 $PP$ decays discussed in Refs. \cite{chang2,chang3}, which provides further
 evidence to support the speculation regarding the topology-dependent
 annihilation parameters reported in Ref. \cite{zhu1,zhu2};
 (iii) 
 Interestingly,  the spaces of (${\rho}_{H}$, ${\phi}_{H}$) 
 (green points) are significantly separated from those of (${\rho}^{f}_{A}$, ${\phi}^{f}_{A}$) but overlap partly with the regions of 
 (${\rho}^{i}_{A}$, ${\phi}^{i}_{A}$), which implies 
 that the simplification $X_{H}$ ${\simeq}$ $X_{A}^{i}$ is
 roughly allowed for $B$ ${\to}$ $PV$ decays as in the case of $B$ ${\to}$ $PP$ decays~\cite{chang3}. 

  %%%%%%%%%%%%%%%%%%%%%%%%%%%%%%%%%%%%%%%%
  \begin{figure}[htbp]
  \subfigure[]{\includegraphics[width=6cm]{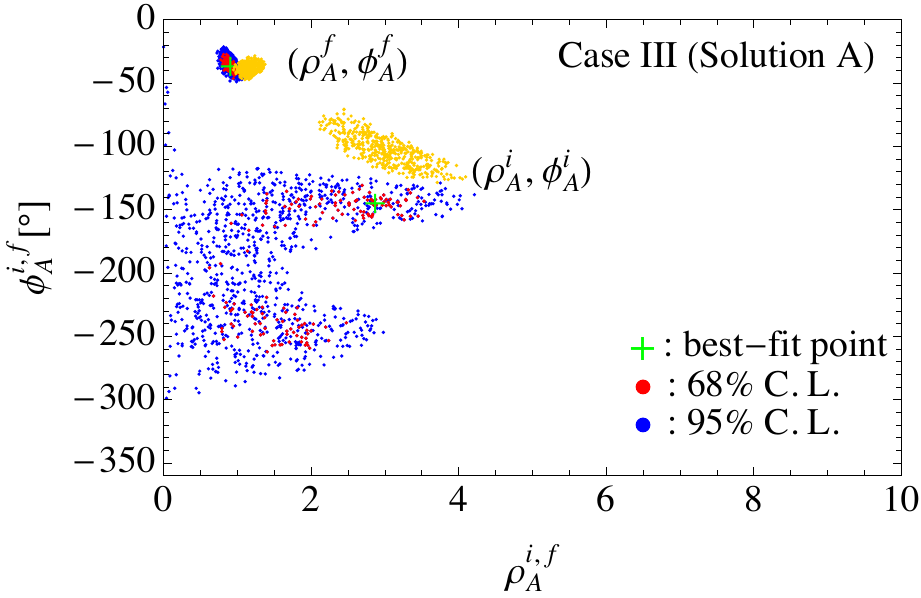}} \quad
  \subfigure[]{\includegraphics[width=6cm]{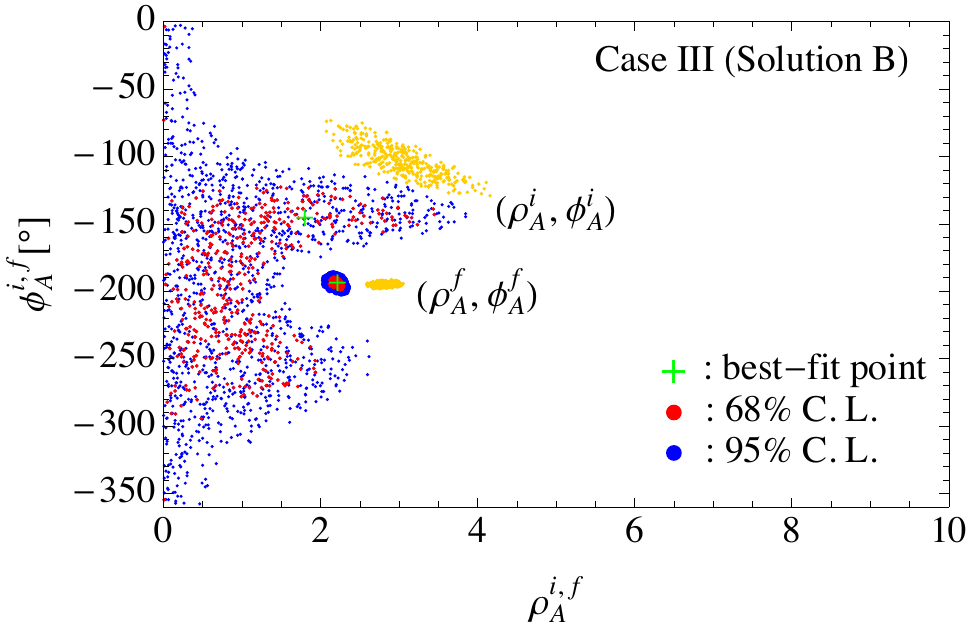}}
  \caption{The allowed regions of parameters 
  (${\rho}_{A}^{i,f}$, ${\phi}_{A}^{i,f}$) at 68\% C.L. 
  and 95\% C.L. indicated by red and blue points, respectively.
  The best-fit points of Solutions A and B correspond to 
  ${\chi}^{2}_{\rm min}$ $=$ $23$ and $26$, respectively.
  For comparison, the fitted results for $B$ ${\to}$ $PP$
  decays \cite{chang3} at 68\% C.L. are also indicated
  by yellow points.}
  \label{figCaseIII}
  \end{figure}
  %%%%%%%%%%%%%%%%%%%%%%%%%%%%%%%%%%%%%%%%

 (3) For Case III, based on the above-described analysis, we will present the most 
 simplified scenario with four free parameters, i.e., (${\rho}^{f}_{A}$, ${\phi}^{f}_{A}$)
 and (${\rho}^{i}_{A}$, ${\phi}^{i}_{A}$) $=$ (${\rho}_{H}$, ${\phi}_{H}$).
 Combining the constraints from 35 independent observables of $B$ ${\to}$ ${\pi}K^{\ast}$, 
 ${\rho}K$, ${\pi}{\rho}$, ${\phi}K$ decays, our fitted results
 are shown in Fig. \ref{figCaseIII}, 
 where ``solution A and B'' corresponds
 to the minimal values ${\chi}^{2}_{\rm min}$ $=$ $23$ and $26$, 
 respectively. Strictly speaking, solution A should be favored over 
 solution B because ${\chi}^{2}_{\rm min,A}$ $<$ ${\chi}^{2}_{\rm min,B}$. For solution A, the allowed spaces of (${\rho}^{i}_{A}$, ${\phi}^{i}_{A}$) at $68\%$ C.L. consist of two separate parts located on two sides of ${\rho}^{i}_{A}=-180^{\circ}$. Corresponding to the best-fit point of solution A, the numerical results of the end-point parameters are     
 \begin{equation}
 ({\rho}_{A,H}^{i},{\phi}_{A,H}^{i}[^{\circ}]) = (2.87^{+0.66}_{-1.95}, -145^{+14}_{-21}), \quad
 ({\rho}_{A}^{f},{\phi}_{A}^{f}[^{\circ}]) = (0.91^{+0.12}_{-0.13}, -37^{+10}_{-9})\,.
   \label{SPVA} \\
\end{equation}
 
 From Fig. \ref{figCaseIII}, it is observed that
 (i) Similarly to Case II, the parameters (${\rho}^{f}_{A}$, ${\phi}^{f}_{A}$)
 are severely restricted to two small and tight spaces.
 (ii) In contrast with Case II, the allowed regions of parameters 
 (${\rho}^{i}_{A}$, ${\phi}^{i}_{A}$) at 68\% C.L. shrink notably 
 due to the simplification $X_{H}$ $=$ $X_{A}^{i}$.
 (iii) The allowed regions of parameters (${\rho}^{f}_{A}$, ${\phi}^{f}_{A}$)
 are completely separated from those of (${\rho}^{i}_{A}$, ${\phi}^{i}_{A}$),
 which implies that the factorizable annihilation parameters $X_{A}^{f}$ should be different from the
 nonfactorizable annihilation parameters $X_{A}^{i}$.
 (iv) The spaces of (${\rho}^{i,f}_{A}$, ${\phi}^{i,f}_{A}$) for $B$ ${\to}$
 $PV$ decays are separated from the spaces of $B$ ${\to}$ $PP$ decays (yellow 
 points in Fig. \ref{figCaseIII}), which implies that parameters $X_{A}$
 for $B$ ${\to}$ $PP$ and $PV$ decays should be introduced 
 and treated individually. 
  
 %%%%%%%%%%%%%%%%%%%%%%%%%%%%%%%%%%%%%%%%
 \begin{table}[ht]
 \caption{The $CP$-averaged branching ratios (in units of $10^{-6}$) and
  direct $CP$ asymmetries  (in units of $10^{-2}$) of $B$ ${\to}$ ${\pi}K^{\ast}$, ${\rho}K$, 
  ${\pi}{\rho}$ and $KK^{\ast}$ decays. For the theoretical results of Case III,
  the first and the second theoretical errors are caused by the 
  CKM parameters and the other parameters (including the quark masses, decay
  constants, form factors and $\lambda_B$), respectively.}
 \label{brcp}
 \begin{ruledtabular}
 \begin{tabular}{c|ccc|ccc}
   Decay 
   & \multicolumn{3}{c|}{Branching fractions}
   & \multicolumn{3}{c}{Direct $CP$ asymmetries}  \\ \cline{2-7}
   modes 
   & Exp. & Case III & S4 & Exp. & Case III & S4 \\ \hline
 $B^{-}$ ${\to}$ ${\pi}^{-}\bar{K}^{{\ast}0}$
   & $10.5\pm0.8$    & $8.7^{+0.4+1.3}_{-0.5-1.2}$        & $8.4$
   & $-4.2{\pm}4.1$ & $0.47^{+0.02+0.11}_{-0.02-0.13}$ & $0.8$ \\
 $B^{-}{\to}{\pi}^{0}K^{{\ast}-}$
   & $8.8{\pm}1.2$  & $5.4^{+0.3+0.7}_{-0.3-0.7}$        & $6.5$
   & $ -39{\pm}12$  & $0.4^{+0.0+4.0}_{-0.0-4.7}$        & $-6.5$ \\
 $\bar{B}^{0}$ ${\to}$ ${\pi}^{+}K^{\ast-}$
   & $8.4{\pm}0.8$       & $7.5^{+0.4+1.1}_{-0.5-1.0}$        & $8.1$
   & $-23{\pm}6$         & $-26^{+1+1}_{-1-1}$                & $-12.1$ \\
 $\bar{B}^{0}$ ${\to}$ ${\pi}^{0}\bar{K}^{{\ast}0}$   
   & $3.3{\pm}0.6$       & $2.9^{+0.1+0.5}_{-0.2-0.5}$        & $2.5$
   & $-15{\pm}13$       & $-21^{+1+6}_{-1-6}$                 & $1.0$ \\ \hline
 $B^{-}$ ${\to}$ $\bar{K}^{0}{\rho}^{-}$
   & $9.4^{+1.9}_{-3.2}$ & $7.9^{+0.4+1.3}_{-0.5-1.1}$        & $9.7$
   & $21^{+31}_{-28}$        & $1.3^{+0.1+0.1}_{-0.1-0.1}$        & $0.8$ \\
 $B^{-}$ ${\to}$ $K^{-}{\rho}^{0}$
   & $3.74^{+0.49}_{-0.45}$  & $3.41^{+0.19+0.63}_{-0.21-0.57}$ & $4.3$
   & $37{\pm}11$              & $26^{+1+5}_{-1-5}$                 & $31.7$ \\
 $\bar{B}^{0}$ ${\to}$ $K^{-}{\rho}^{+}$
   & $7.0{\pm}0.9$            & $9.0^{+0.5+1.4}_{-0.5-1.3}$    & $10.1$
   & $20{\pm}11$              & $27^{+1+3}_{-1-3}$             & $20$ \\
  $\bar{B}^{0}$ ${\to}$ $\bar{K}^{0}{\rho}^{0}$ 
   & $4.7{\pm}0.7$           & $5.5^{+0.3+0.8}_{-0.3-0.7}$      & $6.2$
   & $6{\pm}20$              & $15^{+1+3}_{-1-3}$               & $-2.8$ \\ \hline
 $B^{-}$ ${\to}$ ${\pi}^{-}{\rho}^{0}$
   & $8.3^{+1.2}_{-1.3}$     & $6.8^{+0.6+1.2}_{-0.6-1.1}$      & $12.3$
   & $18^{+9}_{-17}$        & $-6.7^{+0.2+3.2}_{-0.2-3.7}$     & $-11.0$ \\
 $B^{-}$ ${\to}$ ${\pi}^{0}{\rho}^{-}$
   & $10.9^{+1.4}_{-1.5}$  & $10.9^{+0.8+2.7}_{-0.8-2.4}$ & $10.3$
   & $2{\pm}11$            & $8.2^{+0.2+1.6}_{-0.3-1.5}$   & $9.9$ \\
  $\bar{B}^{0}$ ${\to}$ ${\pi}^{+}{\rho}^{-}$ $+$ c.c. % ${\pi}^{-}{\rho}^{+}$
   & $23.0{\pm}2.3$         & $26.7^{+2.1+5.1}_{-2.2-4.5}$ & $23.6$ 
   & --- & --- & --- \\
  $\bar{B}^{0}$ ${\to}$ ${\pi}^{0}{\rho}^{0}$
   & $2.0{\pm}0.5$          & $1.2^{+0.1+0.5}_{-0.1-0.5}$   & $1.1$
   & $-27{\pm}24$          & $-3.9^{+0.1+5.0}_{-0.1-5.1}$   & $10.7$ \\ \hline
 $B^{-}$ ${\to}$ $K^{-}{\phi}$
   & $8.8{\pm}0.5$         & $9.9^{+0.5+1.6}_{-0.6-1.5}$         & $11.6$
   & $4.1{\pm}2.0$         & $0.72^{+0.02+0.14}_{-0.03-0.16}$  & $0.7$  \\
 $\bar{B}^{0}$ ${\to}$ $\bar{K}^{0}{\phi}$
   & $7.3^{+0.7}_{-0.6}$   & $9.3^{+0.4+1.5}_{-0.5-1.4}$ & $10.5$
   & $-1{\pm}14$           & $1.2^{+0.0+0.1}_{-0.0-0.1}$ & $0.8$ \\ \hline 
 $B^{-}$ ${\to}$ $K^{-}K^{{\ast}0}$
   & $<1.1$ & $0.58^{+0.03+0.09}_{-0.04-0.09}$ & $0.66$
   & --- & $-10.6^{+0.3+3.0}_{-0.4-2.6}$     &  $-9.6$ \\
 $B^{-}$ ${\to}$ $K^{{\ast}-}K^{0}$
   & --- & $0.46^{+0.02+0.08}_{-0.03-0.07}$ & $0.55$
   & --- & $-23.0^{+0.6+2.1}_{-0.8-2.2}$      & $-21.1$ \\
 $\bar{B}^{0}$ ${\to}$ $K^{+}K^{{\ast}-}$ $+$ c.c. % $K^{-}K^{{\ast}+}$
   & $<0.4$ & $0.11^{+0.01+0.01}_{-0.01-0.01}$ & $0.15$
   & --- & --- & --- \\
 $\bar{B}^{0}$ ${\to}$ $K^{0}\bar{K}^{{\ast}0}$ $+$ c.c. % $\bar{K}^{0}K^{{\ast}0}$
   & $<1.9$ & $0.96^{+0.05+0.13}_{-0.06-0.11}$ & $1.10$ &--- &--- &--- 
  \end{tabular}
  \end{ruledtabular}
  \footnotetext{Note: Here we adopt the same definition of direct $CP$ asymmetry as HFAG \cite{HFAG}.}
  \end{table}
 %%%%%%%%%%%%%%%%%%%%%%%%%%%%%%%%%%%%%%%%%%%%

 %%%%%%%%%%%%%%%%%%%%%%%%%%%%%%%%%%%%%%%%%%%%
 \begin{table}[ht]
 \caption{The mixing-induced $CP$ asymmetries (in units of $10^{-2}$). 
  The explanation for the uncertainties is the same as that indicated in Table \ref{brcp}.}
 \label{mcp}
 \begin{ruledtabular}
 \begin{tabular}{lccc}
 \multicolumn{1}{c}{Decay modes} & Exp. & Case III  & S4 \\ \hline
  $\bar{B}^{0}$ ${\to}$ $\bar{K}^{0}{\rho}^{0}$ 
  & $54^{+18}_{-21}$  & $63^{+2+3}_{-2-2}$  & --- \\
  $\bar{B}^{0}$ ${\to}$ ${\pi}^{0}{\rho}^{0}$
  & $-23{\pm}34$      & $-29^{+5+3}_{-7-5}$  & --- \\
  $\bar{B}^{0}$ ${\to}$ $\bar{K}^{0}{\phi}$
  & $74^{+11}_{-13}$ & $72^{+2+0}_{-2-0}$   & --- 
 \end{tabular}
 \end{ruledtabular}
 \footnotetext{Note: Here we adopt the same definition of mixing-induced $CP$ asymmetries as HFAG \cite{HFAG}.}
 \end{table}
 %%%%%%%%%%%%%%%%%%%%%%%%%%%%%%%%%%%%%%%%%%%%

 %%%%%%%%%%%%%%%%%%%%%%%%%%%%%%%%%%%%%%%%%%%%
 \begin{table}[ht]
 \caption{The $CP$ asymmetry parameters (in units of $10^{-2}$).
 The explanation for the uncertainties is the same as that indicated in Table \ref{brcp}.}
  \label{cpp}
  \begin{ruledtabular}
  \begin{tabular}{c|ccc|ccc|ccc}
  $CP$ asymmetry 
  & \multicolumn{3}{c|}{$\bar{B}^{0}$ ${\to}$ ${\pi}^{+}{\rho}^{-}$ $+$ c.c.} % ${\pi}^{-}{\rho}^{+}$} 
  & \multicolumn{3}{c|}{$\bar{B}^{0}$ ${\to}$ $K^{+}K^{{\ast}-}$ $+$ c.c.} % $K^{-}K^{{\ast}+}$}
  & \multicolumn{3}{c}{$\bar{B}^{0}$ ${\to}$ $K^{0}\bar{K}^{{\ast}0}$ $+$ c.c.} % $\bar{K}^{0}K^{{\ast}0}$} 
  \\ \cline{2-10}
  parameters & Exp. & Case III & S4 & Exp. & Case III & S4 & Exp. & Case III & S4 \\ \hline
  $C$ & $-3{\pm}6$ & $4.6^{+0.2+0.8}_{-0.2-0.9}$ & $5$ 
       & --- & $0^{+0+0}_{-0-0}$             & ---
       & --- & $13.0^{+0.5+0.6}_{-0.4-0.7}$ & --- \\
 $S$  & $6{\pm}7$ & $-3.6^{+5.0+1.6}_{-6.8-1.6}$ &$9$
       & --- & $12^{+5+0}_{-7-0}$           & ---
       & --- & $4.2^{+0.2+0.7}_{-0.1-0.7}$  & --- \\
 ${\Delta}C$ & $27{\pm}6$ & $33^{+1+14}_{-1-15}$  & $0$
       & --- & $0^{+0+0}_{-0-0}$             & ---
       & --- & $-15.3^{+0.1+9.1}_{-0.1-8.7}$ & --- \\
 ${\Delta}S$ &$1{\pm}8$ &$-1.8^{+0.2+0.9}_{-0.3-0.8}$ & $-3$
       & --- & $0^{+0+0}_{-0-0}$              & ---
       & --- & $-25.0^{+0.3+6.4}_{-0.2-5.7}$ & --- \\
  $A_{CP}$ &$-11{\pm}3$ & $-11.8^{+0.4+1.5}_{-0.3-1.7}$ & $-8$
       & --- & $0^{+0+0}_{-0-0}$             & ---
       & --- & $-10.7^{+0.3+2.2}_{-0.4-2.1}$ & ---
 \end{tabular}
 \end{ruledtabular}
 \footnotetext{Note: Here we adopt the same definition for the parameters 
  $C_{f\bar{f}}$, $S_{f\bar{f}}$, ${\Delta}C_{f\bar{f}}$, ${\Delta}S_{f\bar{f}}$
  and $A_{CP}^{f\bar{f}}$ as HFAG \cite{HFAG} and choose the final states $f$ $=$ 
  ${\rho}^{+}{\pi}^{-}$, $K^{{\ast}+}K^{-}$ and $K^{{\ast}0}\bar{K}^{0}$.}
 \end{table}
%%%%%%%%%%%%%%%%%%%%%%%%%%%%%%%%%%%%%%%%%%%%

 Using the best-fit~(central) values of solution A in Eq.(\ref{SPVA}), we present the 
 theoretical results for the branching fractions and $CP$ asymmetries 
 of $B$ ${\to}$ $PV$ decays in the ``Case III'' columns of 
 Tables \ref{brcp}, \ref{mcp} and \ref{cpp}.   
 For comparison, the theoretical results of ``Scenario S4'' \cite{Beneke2}, with 
  (${\rho}_{A}^{PV}$, ${\rho}_{A}^{PV}$) $=$ (1, $-20^{\circ}$) and
  (${\rho}_{A}^{VP}$, ${\rho}_{A}^{VP}$) $=$ (1, $-70^{\circ}$), are
  also listed in the ``S4'' columns of the tables.
 It is observed that most of our theoretical results are consistent with 
 the experimental data
 except for a few contradictions in the $B^{-}{\to}{\pi}^{0}K^{{\ast}-}$ decay, which will be discussed later, and are similar to the ``S4'' results.
 
For the well-measured observables, such as the branching ratios
 ${\cal B}(B{\to}{\phi}K)$, ${\cal B}(B^{-}{\to}{\pi}^{-}{\rho}^{0})$, 
 ${\cal B}(B^{0}{\to}K{\rho})$ with a significance level ${\ge}$ $6{\sigma}$ (see Table \ref{brcp}), the direct $CP$ asymmetry for 
 $\bar{B}^{0}$ ${\to}$ ${\pi}^{+}K^{{\ast}-}$ decay 
 (see Table \ref{brcp}) and ${\Delta}C$ for $B$ ${\to}$ 
 ${\pi}^{\pm}{\rho}^{\mp}$ decay (see Table \ref{cpp}) 
 with a significance level ${\ge}$ $4{\sigma}$,  compared with the traditional ``S4'' results, 
 our results are more in line with the experimental data.
  In particular, compared with the measurement
 ${\Delta}C$ $=$ $(27{\pm}6)\%$ for $B$ ${\to}$ 
 ${\pi}^{\pm}{\rho}^{\mp}$ decay, the difference 
 between the ``S4'' results and ours is clear and 
 notable, which may imply that a relatively large 
 ${\rho}_{A,H}^{i}$ ${\sim}$ $3$ rather than the conventionally
 used small ${\rho}_{A}^{i}$ ${\sim}$ 1 \cite{Beneke2,Cheng3,Cheng1,Cheng2}
 may be necessary for nonfactorizable annihilation corrections.
 In addition, evidence of a large ${\rho}_{A}$ for $B$ ${\to}$ $K{\rho}$,
 $K^{\ast}{\pi}$ decays is also presented in Fig. 3 of Ref.\cite{1409.3252} using a
 similar ${\chi}^{2}$ fit approach, with the simplification that $X_{A}^{i}$ $=$ $X_{A}^{f}$.

 %%%%%%%%%%%%%%%%%%%%%%%%%%%%%%%%%%%%%%%%
  \begin{figure}[htbp]
  \includegraphics[width=6cm]{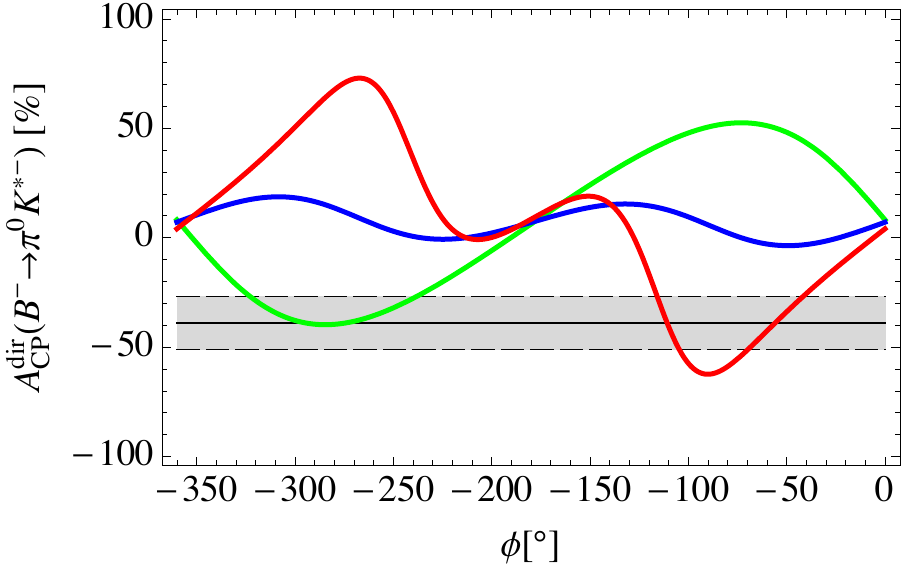}
  \caption{The green, blue and red lines correspond to the dependence of $A^{dir}_{CP}(B^{-}{\to}{\pi}^{0}K^{{\ast}-})$ on $\phi_H$, ${\phi}_{A}^{i}$ and ${\phi}_{A}^{f}$, with $\rho_H=3~(\rho_A^{i,f}=0)$, ${\rho}_{A}^{i}=3~(\rho_{A,H}^{f}=0)$ and ${\rho}_{A}^{f}=1~(\rho_{A,H}^{i}=0)$, respectively. The shaded region corresponds to experimental data~($1\sigma$ error bar). }
  \label{figdep}
  \end{figure}
  %%%%%%%%%%%%%%%%%%%%%%%%%%%%%%%%%%%%%%%%

Unfortunately, with the central values presented in Eq.~(\ref{SPVA}), from the results gathered in Table~\ref{brcp}, one may find that our result $A^{dir}_{CP}(B^{-}{\to}{\pi}^{0}K^{{\ast}-})=(0.4^{+0.0+4.0}_{-0.0-4.7})\%$ is significantly larger than the data $(-39\pm12)\%$ reported by BABAR. To clarify the reason for this discrepancy, we present the dependence of $A^{dir}_{CP}(B^{-}{\to}{\pi}^{0}K^{{\ast}-})$ on $\phi_H$, ${\phi}_{A}^{i}$ and ${\phi}_{A}^{f}$ in Fig.~\ref{figdep}. It is easily observed that the best-fit result $({\rho}_{A}^{f},{\phi}_{A}^{f})\sim (0.91, -37^{\circ})$ is favored by the BABAR data. However, the best-fit value $({\rho}_{A,H}^{i},{\phi}_{A,H}^{i})$ $\sim (2.87, -145^{\circ})$ results in the large mismatch for $A^{dir}_{CP}(B^{-}{\to}{\pi}^{0}K^{{\ast}-})$~(in Eq.(\ref{SPVA}), a small ${\rho}_{H}$ is also allowed at $68\%$ C.L., which would yield a better agreement but result in a relative larger $\chi^2$).
One interesting and important problem is that 
 a relatively large ${\rho}_{H}$ ${\sim}$ $3$ in $B$ ${\to}$ $PP$ decays, which is similar to the best-fit value for $B$ ${\to}$ $PV$ decays in this work, 
 is always required to enhance ${\alpha}_{2}$ contributions
 in resolving the ``${\pi}{\pi}$'' and ``${\pi} K$"puzzles~\cite{chang2,chang3} but clearly leads to a wrong sign for $A^{dir}_{CP}(B^{-}{\to}{\pi}^{0}K^{{\ast}-})$ when confronted with BABAR data, as indicated herein and in Ref.~\cite{Cheng4}. Therefore, if a large negative $A^{dir}_{CP}(B^{-}{\to}{\pi}^{0}K^{{\ast}-})$ is confirmed by Belle~(or future Belle~II) and LHCb collaborations, resolving the ``${\pi}{\pi}$'' and ``${\pi} K$" puzzles through color-suppressed tree amplitude ${\alpha}_{2}$ will be challenging. If so, a large complex electroweak amplitude $\alpha_{3,EW}$ is probably required~\cite{Cheng4}, which may hint possible new physics effects. In addition, the measurements for observables of $B_s\to\phi\pi^0$ decay, whose amplitude is related to ${\alpha}_{2}$ and $\alpha_{3,EW}$ only, may provide a clue even though such decay mode is not easily to be measured soon.
  
 For the color-suppressed tree-dominated $B$ ${\to}$ ${\pi}^{0}{\rho}^{0}$ decay, 
 the penguin-dominated $B^{-}$ ${\to}$ $KK^{\ast}$ decays and
 the pure annihilation $\bar{B}^{0}$ ${\to}$ $K^{\pm}K^{{\ast}{\mp}}$ decay,
 the decay amplitudes are sensitive to the nonfactorizable HSS and 
 annihilation corrections, and their measurements could perform strong 
 constraints on $X_{A,H}^{i}$(${\rho}^{i}_{A,H}$, ${\phi}^{i}_{A,H}$).
 Unfortunately, the experimental errors of the observables for $B$ ${\to}$
 ${\pi}^{0}{\rho}^{0}$ decay are too large, and the $B^{-}$ ${\to}$ $KK^{\ast}$ and $\bar{B}^{0}$ ${\to}$
 $K^{\pm}K^{{\ast}{\mp}}$ decays have not yet been observed. 
 Future refined measurements conducted at the LHCb and Belle II
 would be very helpful in carefully examining the HSS and 
 annihilation corrections.
 Recently, the LHCb collaboration has updated the upper limit of branching fractions
 for pure annihilation $\bar{B}^{0}$ ${\to}$ $K^{\pm}K^{{\ast}{\mp}}$ decay
 with $<$ $0.4(0.5){\times}10^{-6}$ at 90 (95\%) C.L. \cite{LHCbKKV}, and 
 it is eagerly expected that these decays can be precisely measured, which should be useful in probing the annihilation corrections and the corresponding mechanism.
 Of course, one can use different mechanisms for enhancing
 the nonfactorizable contributions in QCDF, for example, the final
 state rescattering effects advocated in Ref. \cite{Cheng3,Cheng1,Cheng2},
 in which the allowed regions for parameters (${\rho}^{i}_{A,H}$, ${\phi}^{i}_{A,H}$)
 might be different.

 In summary, we studied the contributions of HSS and annihilation in
 $B$ ${\to}$ $PV$ decays within the QCDF framework. 
 Unlike the traditional treatment of annihilation endpoint divergence
 with process-dependent parameters (${\rho}_{A}^{PV}$, ${\phi}_{A}^{PV}$) and 
 (${\rho}_{A}^{VP}$, ${\phi}_{A}^{VP}$) in previous studies
 \cite{Beneke2,Cheng3,Cheng1,Cheng2},
 the topology-dependent parameters (${\rho}_{A}^{i,f}$, ${\phi}_{A}^{i,f}$)
 based on a recent analysis of $B$ ${\to}$ $PP$ decays
 \cite{zhu1,zhu2,chang2,chang3} were used in this paper.
 Combining available experimental data, we performed comprehensive $\chi^2$ 
 analyses of $B$ ${\to}$ $PV$ decays and obtained information
 and constraints regarding the parameters (${\rho}_{A}^{i,f}$, ${\phi}_{A}^{i,f}$).
 It is observed that most of the measurements on observables of $B$ ${\to}$ $PV$
 decays, except for some contradictions in $B^{-}{\to}{\pi}^{0}K^{{\ast}-}$ decay, could be properly interpreted with the best-fit values presented in Eq.(\ref{SPVA}),
 which suggests that the topology-dependent parameterization of
 annihilation and HSS corrections may be suitable.
 The other findings of this study are summarized as follows: 
 \begin{itemize}
 \item
 The relatively small value of the $B$ wave function parameter 
 ${\lambda}_{B}$ ${\sim}$ 0.2 GeV, which is only related to 
 the universal $B$ wave functions and plays an important role
 in providing a possible solution to the so-called ``${\pi}{\pi}$'' and
 ``${\pi}K$'' puzzles \cite{chang2}, is also allowed by the constraints from $B$
 ${\to}$ $PV$ decays.
 \item
 As used extensively in phenomenological studies on hadronic 
 $B$ decays \cite{Beneke2,Cheng3,Cheng1,Cheng2}, generally,
 parameters $X_{A,H}^{i,f}$ for $B$ ${\to}$ $PP$ 
 and $PV$ decays should be independent 
 of each other and be treated individually.
 \item
 The allowed regions of parameters (${\rho}_{A}^{f}$, ${\phi}_{A}^{f}$)
 are strictly constrained by available experimental data, whereas the accessible 
 spaces of parameters (${\rho}_{A}^{i}$, ${\phi}_{A}^{i}$) are relatively
 large. 
 Generally, there is no common space between 
 (${\rho}_{A}^{f}$, ${\phi}_{A}^{f}$) and 
 (${\rho}_{A}^{i}$, ${\phi}_{A}^{i}$) with the approximation of
 $X_{A}^{i}$ $=$ $X_{H}$, which implies that factorizable
 annihilation parameters $X_{A}^{f}$ should be different from
 nonfactorizable annihilation parameters $X_{A}^{i}$. 
 Moreover, a relatively large ${\rho}_{A}^{i}$ ${\sim}$ 3
 is required by the considerable fine-tuning of $X_{A}^{i,f}$ 
 to reproduce most of the measurements on hadronic $B$ decays.
 The above-described evidence and features have been clearly observed
 in both $B$ ${\to}$ $PP$ decays \cite{zhu1,zhu2,chang2,chang3}
 and $B$ ${\to}$ $PV$ decays.
  \item Unfortunately, a relatively large ${\rho}_{H}$ ${\sim}$ $3$ with $\phi_H\sim -145^{\circ}$ related to significant HSS corrections to color-suppressed tree amplitude $\alpha_2$, which is helpful for resolving the ``${\pi}{\pi}$'' and ``${\pi} K$" puzzles and allowed by most $B\to PP$ and $PV$ decays, result in a wrong sign for $A^{dir}_{CP}(B^{-}{\to}{\pi}^{0}K^{{\ast}-})$ when confronted with recent BABAR data $(-39{\pm}12)\%$. This finding suggests a large, complex electroweak amplitude attributed to possibly new physics or an undiscovered mechanism~\cite{Cheng4}, which deserves much attention. Before we know for sure, the crosscheck based on refined measurements conducted at Belle (Belle II) and LHCb is urgently awaited.
 \end{itemize}
 
 Overall, the annihilation and HSS contributions in nonleptonic $B$
 decays should be and have been attracting much attention and careful
 study. For $B$ ${\to}$ $PV$ decays, a comparative advantage
 is that there are more decay modes and more observables than 
 those for $B$ ${\to}$ $PP$ decays, and hence more information
 and more stringent constraints on parameters $X_{A,H}$
 can be obtained, which represents an opportunity as well
 as a challenge in the rapid accumulation of data on $B$
 events at running LHCb and forthcoming Belle II/SuperKEKB.
 Theoretically, these results will surely help us to further understand the underlying
 mechanism of annihilation and HSS contributions
 and develop more efficient approaches to calculate hadronic
 matrix elements. 

 \section*{Acknowledgments}
 This work is supported by the National Natural Science Foundation of China
 (Grant Nos.  11475055, 11105043, 11147008, 11275057 and U1232101).
 Q. Chang is also supported by the Foundation for the Author of 
 National Excellent Doctoral Dissertation of P. R. China (Grant No. 201317)
 and the Program for Science and Technology Innovation Talents in
 Universities of Henan Province (Grant No. 14HASTIT036). We also thank the Referee and Hai-Yang Cheng for their helpful comments.

 \begin{appendix}
 \section*{Appendix: Theoretical input parameters}
 For the CKM matrix elements using the Wolfenstein parameterization,
 we adopt the fitting results given by the CKMfitter group \cite{CKMfitter}
 \[
 \bar{\rho}=0.1453^{+0.0133}_{-0.0073}, 
 \quad
 \bar{\eta}=0.343^{+0.011}_{-0.012},
 \quad
 A=0.810^{+0.018}_{-0.024},
 \quad
 \lambda=0.22548^{+0.00068}_{-0.00034}.
 \]

 The pole and running masses of quarks used in our analysis are \cite{PDG14}
\[
 m_{u,d,s}=0,
 \quad
 m_{c}=1.67{\pm}0.07\,{\rm GeV},
 \quad
 m_{b}=4.78{\pm}0.06\,{\rm GeV},
 \]
 \[
 \frac{\bar{m}_s(\mu)}{\bar{m}_q(\mu)}=27.5{\pm}1.0,
 \quad
 \bar{m}_{s}(2\,{\rm GeV})=95{\pm}5\,{\rm MeV},
 \quad
 \bar{m}_{b}(\bar{m}_{b})=4.18{\pm}0.03\,{\rm GeV},
 \]
 where $m_q$ $=$ $m_{u}$ $=$ $m_{d}$ $=$ $(m_{u}+m_{d})/2$. 

 The decay constants of pseudoscalar and vector mesons are \cite{PDG14,DecayCon,PBall2007}
 \begin{eqnarray}
 &f_{B}=(190.6{\pm}4.7)\,{\rm MeV},
  \quad
  f_{\pi}=(130.41{\pm}0.20)\,{\rm MeV},
  \quad
  f_{K}=(156.2{\pm}0.7)\,{\rm MeV},
  \nonumber \\
 &f_{\rho}=(216{\pm}3)\,{\rm MeV},
   \qquad
   f_{\rho}^{\bot}({\rm 1\,GeV})=(165{\pm}9)\,{\rm MeV},
   \nonumber \\
 &f_{K^{\ast}}=(220{\pm}5)\,{\rm MeV},
   \qquad
   f_{K^{\ast}}^{\bot}({\rm 1\,GeV})=(185{\pm}10)\,{\rm MeV}.
   \nonumber
   \end{eqnarray}

  The heavy-to-light transition form factors are \cite{BallZwicky}
  \begin{eqnarray}
 & F^{B{\to}{\pi}}_{1}=0.258{\pm}0.031,
   \quad
   F^{B{\to}K}_{1}=0.331{\pm}0.041,
   \nonumber \\
 & A^{B{\to}{\rho}}_{0}=0.303{\pm}0.029,
   \quad
   A^{B{\to}K^{\ast}}_{0}=0.374{\pm}0.034.
   \nonumber
   \end{eqnarray}

  The Gegenbauer moments are \cite{BallG}
  \begin{eqnarray}
 & a_{1}^{\pi}=0,
   \quad\ \
   a_2^{\pi}({\rm 1\,GeV}) =0.25,
   \quad\ \
   a_{1}^{K}({\rm 1\,GeV}) =0.06,
   \quad\ \
   a_{2}^{K}({\rm 1\,GeV}) =0.25,
   \nonumber \\
 & a_{1,\rho}^{||}=0,
   \quad
   a_{2,\rho}^{||}({\rm 1\,GeV})=0.15,
   \quad
   a_{1,K^{\ast}}^{||}({\rm 1\,GeV})=0.03,
   \quad
   a_{2,K^{\ast}}^{||}({\rm 1\,GeV})=0.11.
   \nonumber
   \end{eqnarray}
   
   For other inputs, such as the masses and lifetimes of mesons 
   {\it et al.}, we adopt the values given by PDG \cite{PDG14}.
  \end{appendix}

 \end{document}